# Microwave Radiation from a Particle Revolving Along a Shifted Equatorial Orbit About a Dielectric Ball

Levon Sh. Grigoryan, Hrant F. Khachatryan, Svetlana R. Arzumanyan, and Mher L. Grigoryan
Institute of Applied Problems in Physics, Yerevan, 0014 Armenia

*Abstract*—A relativistic electron uniformly rotating along a shifted equatorial orbit about a dielectric ball may generate microwave Cherenkov radiation tens of times more intense as that generated at the revolution in a continuous, infinite and transparent medium.

## I. INTRODUCTION AND BACKGROUND

THE operation of a number of devices intended for production of electromagnetic radiation is based on the interaction of relativistic electrons with matter [1,2]. Numerous applications motivate the importance of exploring different mechanisms of amplification and control of produced radiation. Specifically, one may use radiation of different kinds as well as reflecting surfaces for monitoring the flows of radiation.

A research of such a kind was carried out in [3,4], where the emission of radiation from a relativistic particle revolving about a dielectric ball has been investigated. In such a geometry in addition to the synchrotron radiation (SR) the particle may generate the Cherenkov radiation (CR), since the field associated with the particle partially penetrates the ball depths and revolves together with the particle. In case of short distances of particle from the ball surface the velocity of its field displacement within the ball may exceed the phase velocity of light in the ball material and, hence, CR would be generated. Here the flow of produced CR may be controlled either by changing the radius of ball or the relative position of ball with respect to particle orbit.

As a result of the combination of SR and CR, as well as of the mechanism of radiation flow control (the ball-to-vacuum interface) the revolving particle may, at separate harmonic, generate CR tens of times more intense than CR and SR from the same particle generated in a continuous, infinite and transparent medium having the same permittivity as the ball material. The theoretical substantiation of this effect and its visual explanation are given in [3].

However, the case under consideration in [3,4] was confined to the simplest configuration of ball located in the centre of particle orbit (Fig.1a). In reality (electron beam), such a symmetry is difficult to provide. In the present work the case of shifted configuration of the system (Fig. 2a) is considered when $d \neq 0$, where $d$ is the distance between the centers of the ball and particle orbit.

Now consider the uniform rotation of a relativistic electron in the magnetic field in vacuum about a dielectric ball in its equatorial plane under the assumption that the centers of ball and of particle orbit are shifted one with respect to the other.

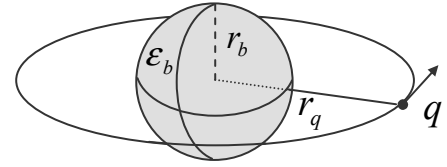
Fig.1a

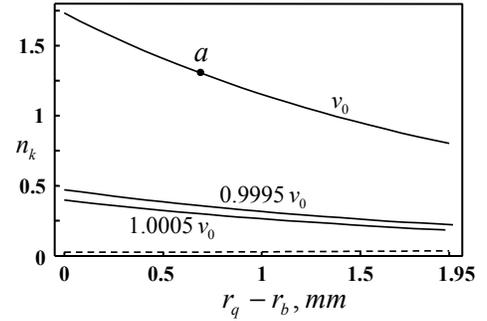
Fig.1b

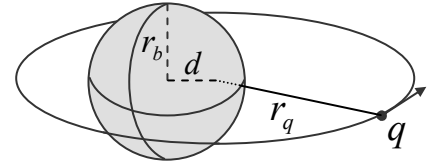
Fig.2a

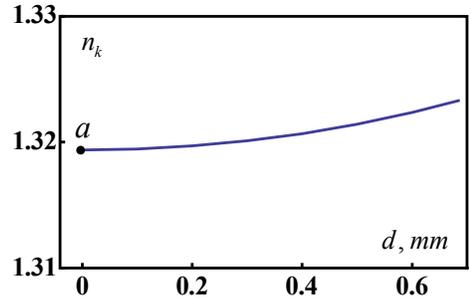
Fig.2b

The rotation of particle entails radiation at some discrete frequencies (harmonics) $\nu_k = k\nu_q$ with $k = 1;2;3...$. We assume that an exterior force would make up for the braking of particle due to the radiation, by forcing the particle to uniformly rotate about the ball. The total energy losses of particle during one revolution period are written as

$$\sum_k (E_k^{(d)} + E_k^{(r)}) = \sum_k W_k , \quad (1)$$

where $E_k^{(d)}, E_k^{(r)}$ are the dielectric losses of energy and energy losses due to particle radiation.

It is convenient to introduce a dimensionless quantity

$$E_k^{(r)}/h\nu_k \equiv n_k, \qquad (2)$$

where $h\nu_k$ is the energy of corresponding electromagnetic wave quantum. So, $n_k$ is the "number of electromagnetic field quanta" emitted during one revolution period of particle.

The expressions for calculation of $W_k(d)$ have been derived in the present paper. $W_k$ for the special case of $d=0$ was calculated in [3].

## II. RESULTS

We shall assume that the ball material is a loss-free dielectric and therefore $W_k = E_k^{(r)}$. The permittivity of the ball material is $\varepsilon_b = 3.78$ (molten quartz in $\sim 10^{10}\,Hz$ frequency range), and its radius is $r_b = 36.2\,mm$. We shall confine ourselves to the consideration of electron radiation at a separate harmonic $k=8$.

First, let us consider the case of $d=0$ [3,4]. Shown in Fig.1b are three plots of the number of photons $n_k$ emitted during one turn of electron. Here the distance of particle to the ball surface, $r_q - r_b$, is an independent variable. The rotation frequency $\nu_q$ of electron along each curve is constant. Appropriate values of $\nu_q$ are specified beside the curves. Here $\nu_0 = 1250.7\,MHz$ [3]. For comparison in Fig.1b we give also the value of $n_k(\infty) = 0.027$ for a continuous, infinite nd transparent medium with $\varepsilon = \varepsilon_b$ (dotted line). In the absence of ball only SR is generated and $n_k(vac) \approx 0.005 \ll n_k(\infty)$. It is seen that as compared with $n_k(\infty)$ tens of times larger values

$$0.8 \leq n_k \leq 1.4 \qquad (3)$$

are possible. Here $r_q - r_b$ and the energy of particle $E_q$ may change within sufficiently wide limits:

$$0.5 \leq r_q - r_b \leq 1.8\,mm, \quad 1.9 \leq E_q \leq 5.8\,MeV. \qquad (4)$$

A comparison of curves with $\nu_q/\nu_0 = 1$; 0.9995 and 1.0005, illustrates how rapidly the function $n_k(\nu_q)$ tends to a local extremum $n_k(\nu_0)$ when $\nu_q \to \nu_0$. To increase $n_k$ to about $\sim 1$ the particle should be rotated at "resonance" frequency $1250.7\,MHz$ with $0.3\,MHz$ error (e.g., in cyclotrons, synchrotrons and electron storage rings the frequency of electron rotation is kept constant with higher accuracy).

Now direct our attention to the case when the centers of the ball and of particle orbit do not coincide. In Fig.2b the dependence of $n_k$ on the distance $d$ between the centers of the ball and particle orbit is shown. The rotation frequency of electron $\nu_q = \nu_0$. Corresponding to the 8-th harmonic is the microwave radiation with the wavelength equal to $3\,cm$ in vacuum. The energy of electron is $E_q = 2\,MeV$, and the radius of its orbit is $r_q = 36.884\,mm$. Points $a$ in Fig.2b and Fig.1b correspond to the same state of the system.

In [3] the value of $n_k(d)$ for $d=0$ was calculated using another method and the obtained value of 0.95 is less than $n_k(0) = 1.32$ shown in Fig. 2b, since in [3] the allowance was made for the absorption of some part of microwave radiation from electron by the material of ball. As is seen in Fig. 2b, the radiation stays intensive when $d$ increases from 0 to maximum $0.684\,mm = r_q - r_b$. Here

$$n_k/n_k(\infty) \approx 50. \qquad (5)$$

The value of $n_k$ for an electron uniformly revolving along a non-equatorial orbit about the dielectric ball was calculated in [5]. It was shown that the relation (5) remains valid for distances of the orbit plane to the centre of ball not exceeding $0.1r_q$.

So, in order that the radiation have parameters (3), (5), the frequency of particle rotation should be given with relatively high precision to $\sim 0.02\%$, whereas the positions of the centers of ball and particle orbit may be given with less accuracy. The last fact simplifies possible applications of such a radiation.